\documentclass[,twocolumn,showpacs,aps,pra,superscriptaddress]{revtex4-2}
\usepackage{amsfonts}
\usepackage{amsmath}
\usepackage{amssymb}
\usepackage{mathrsfs}
\usepackage{extarrows}
\usepackage{bm}
\usepackage{graphicx}
\usepackage{overpic}
\usepackage{color}
\usepackage{xcolor}
\usepackage{csquotes}
\usepackage{natbib}
\usepackage{url}
\usepackage{ulem}
\usepackage[hidelinks]{hyperref}

\newcommand{\be}{\begin{equation}}
\newcommand{\ee}{\end{equation}}
\newcommand{\bea}{\begin{eqnarray}}
\newcommand{\eea}{\end{eqnarray}}

\newcommand{\Eq}[1]{Eq.\,(\ref{#1})}

\newcommand{\la}{\langle}
\newcommand{\ra}{\rangle}

\newcommand{\Ht}{H_{_{\rm T}}}
\newcommand{\Hs}{H_{_{\rm S}}}

\newcommand{\He}{H_{_{\rm E}}}
\newcommand{\Hse}{H_{_{\rm SE}}}

\newcommand{\mL}{\mathcal{L}}

\makeatletter

\begin{document}

\title{Simulating Non-Markovian Open Quantum Dynamics with Neural Quantum States}

\author{Long Cao} \altaffiliation{These authors contributed equally to this work.}
\affiliation{Hefei National Research Center for Physical Sciences at the Microscale, 
University of Science and Technology of China, Hefei, Anhui 230026, China}

\author{Liwei Ge} \altaffiliation{These authors contributed equally to this work.}
\affiliation{Hefei National Research Center for Physical Sciences at the Microscale, 
University of Science and Technology of China, Hefei, Anhui 230026, China}

\author{Daochi Zhang}
\affiliation{Department of Chemistry, Fudan University, Shanghai 200438, China}

\author{Xiang Li}
\affiliation{Hefei National Research Center for Physical Sciences at the Microscale, 
University of Science and Technology of China, Hefei, Anhui 230026, China}

\author{Jialin Pan}
\affiliation{Hefei National Research Center for Physical Sciences at the Microscale, 
University of Science and Technology of China, Hefei, Anhui 230026, China}

\author{Yao Wang}
\affiliation{Hefei National Research Center for Physical Sciences at the Microscale, 
University of Science and Technology of China, Hefei, Anhui 230026, China}

\author{Rui-Xue Xu}
\affiliation{Hefei National Research Center for Physical Sciences at the Microscale, 
University of Science and Technology of China, Hefei, Anhui 230026, China}
\affiliation{Hefei National Laboratory, Hefei, Anhui 230088, China}

\author{YiJing Yan}
\affiliation{Hefei National Research Center for Physical Sciences at the Microscale, 
University of Science and Technology of China, Hefei, Anhui 230026, China}

\author{Xiao Zheng} \email{xzheng@fudan.edu.cn}
\affiliation{Department of Chemistry, Fudan University, Shanghai 200438, China}
\affiliation{Hefei National Laboratory, Hefei, Anhui 230088, China}

\begin{abstract}

Reducing computational scaling for simulating non-Markovian dissipative dynamics using artificial neural networks is both a major focus and formidable challenge in open quantum systems. To enable neural quantum states (NQSs), we encode environmental memory in dissipatons (quasiparticles with characteristic lifetimes), yielding the dissipaton-embedded quantum master equation (DQME). The resulting NQS-DQME framework achieves a compact representation of both many-body correlations and non-Markovian memory. Benchmarking against numerically exact hierarchical equations of motion confirms NQS-DQME maintains comparable accuracy while enhancing scalability and interpretability. This methodology opens new paths to explore non-Markovian open quantum dynamics in previously intractable systems.

\end{abstract}

\date{Submitted on June~18, 2025; updated on November~12, 2025}

\maketitle

%%%Xiao: all references should be checked and modified carefully upon completion of main text

% \section{Introduction}
{\bf Introduction.} 
Many-body open quantum systems (OQSs) have attracted wide attention due to their profound applications across diversified fields
\cite{Engel2007,doi:10.1126/science.1235820,RevModPhys.94.045008}. 
Accurately characterizing quantum correlations is crucial for the studies of 
%understanding and predicting key properties and dynamic phenomena in 
many-body OQSs. For instance, the Kondo correlation between a magnetic molecule (the primary system) and a substrate (the environment) critically determines the electric conductance across the molecule-substrate interface \cite{wilson1975renormalization,doi:10.1126/science.280.5363.567,doi:10.1126/science.1113449,li2020molecular}, exemplifying spatially extended quantum correlations. Similarly, non-Markovian memory effects arising from system-environment coupling \cite{breuer2016colloquium,de2017dynamics,PhysRevLett.133.140403} can lead to complex scaling relations in the dissipative dynamics of OQSs \cite{anders2005real,Din24174120}, demonstrating temporally persistent quantum correlations \cite{PhysRevLett.111.086601,doi:10.1126/science.aay6779,doi:10.1126/science.abg8223,Ding25084114}.

Numerically exact approaches \cite{Tam18030402,Lam193721,PhysRevLett.82.1801,10.1063/1.1647528,PhysRevLett.123.050601,PhysRevLett.88.170407,suess2014hierarchy,PhysRevLett.115.266802,PhysRevLett.130.186301,PhysRevB.87.115115,PhysRevLett.88.256403,RevModPhys.92.011001,10.1063/1.3173823,RevModPhys.93.015008}, including the hierarchical equations of motion (HEOM) \cite{Tan89101,YAN2004216,Jin08234703,PhysRevLett.109.266403}, have been developed to simulate non-Markovian open quantum dynamics. 
However, their applicability remains significantly constrained by the exponential wall inherent to the quantum many-body problem \cite{kohn1999nobel}, as their computational costs typically scale exponentially with both system size and environmental memory complexity. 
To address this challenge, tensor network states (TNSs) 
\cite{A.Klumper_1993,flannigan2022many,jaschke2018one,link2024open,werner2016positive,Sta183721,Shi18174102,Ke22194102,schwarz2018nonequilibrium,PhysRevLett.123.240602,10.1063/5.0140002} have been utilized to compress the representation of quantum correlations. However, while TNSs successfully reduce the required number of 
dynamical variables, the resulting approaches still encounter difficulties when area laws are violated \cite{eisert2010colloquium}.
Particularly, TNSs often require large bond dimensions when representing intermediate states during temporal evolution \cite{bonnes2014dynamical, weimer2021simulation} and strong system-environment correlation \cite{Shi18174102}.

Neural quantum states (NQSs) \cite{doi:10.1126/science.aag2302}, which leverage the power of artificial neural networks, have emerged as a promising method for tackling the exponential wall problem in many-body OQSs \cite{Hermann2023,sharir2020deep,shang2023solving,PhysRevLett.120.240503,donatella2023dynamics,ye2025simulating}. NQSs have demonstrated more expressive power than TNSs \cite{Che18085104,sharir2022neural,lange2024architectures}, e.g., in determining  the ground states \cite{wu2023tensor,deng2017quantum} and time evolutions \cite{doi:10.1126/science.aag2302} of quantum spin systems.
%%%Xiao: add some refs showing NQS is more compact than MPS/TNS
While NQSs have been applied to open quantum systems, simulations have so far been restricted to Markovian environments \cite{PhysRevLett.122.250501,PhysRevLett.122.250502,PhysRevLett.122.250503,mellak2023quantum,PhysRevB.99.214306,mellak2024deep}. Consequently, it is believed that addressing non-Markovian memory effects poses the ``ultimate toughness test'' for simulating open quantum dynamics \cite{Schuld19}. Moreover, the sign problem inherent to fermionic environments \cite{PhysRevLett.123.050601} introduces an additional challenge \cite{RevModPhys.91.045002}.

A recently developed theory \cite{li2024toward} based on the concept of dissipatons (Brownian quasiparticles with complex energies) \cite{Yan14054105} provides an exact description of non-Markovian open quantum dynamics via the dissipaton-embedded quantum master equation (DQME). In this framework, environmental memory is fully encoded in the dissipatons' statistical behavior. Formulated with second quantization, the DQME treats system and environmental degrees of freedom on an equal footing \cite{li2024toward}, making it amenable to a NQS representation of quantum correlations in many-body OQSs.

This Letter aims to establish a framework to integrate the NQS approach with the DQME, and to demonstrate that the unified NQS-DQME approach has the potential to resolve the ``ultimate toughness test'' of simulating non-Markovian dynamics in strongly correlated OQSs.

\begin{figure}[t]
	\includegraphics[width=\columnwidth]{figure1.pdf}
	\caption{Schematic of the fermionic DQME theory, mapping the original OQS (left) to a dissipaton-embedded system (right). Red and blue bars represent the $N_{_{\rm S}}$ system fermion energy levels and $N_{_{\rm E}}$ memory-carrying dissipaton levels, respectively. 
    Broadening of blue bars indicates each dissipaton's decay rate (inverse lifetime). } \label{fig1}
\end{figure}

{\bf Fermionic DQME method.}
The total Hamiltonian of an OQS comprises three components: $\Ht = \Hs + \He + \Hse$, where $\Hs$, $\He$, and $\Hse$ represent the system, environment, and system-environment coupling Hamiltonians, respectively. For a noninteracting fermionic environment linearly coupled to the system, the environmental influence is fully characterized by the hybridization time-correlation function. This function is decomposed into a sum of exponentials, $C^{\sigma}(t) = \sum_{j} \eta_{j}^\sigma e^{-\gamma_{j}^{\sigma}t}$, which defines the dissipatons \cite{hu2010communication,Yan14054105,li2024toward}.
Here, $j$ labels the dissipaton levels, and $\sigma$ denotes the dissipaton charge ($\sigma=+$ for hole-type; $\sigma=-$ for electron-type). The imaginary and real parts of $\gamma_j^\sigma$ correspond to the dissipaton energy and inverse lifetime, respectively, while $\eta^\sigma_j$ quantifies the system-dissipaton coupling strength. Crucially, the long-time behavior of $C^\sigma(t)$, and hence the non-Markovian effects, is dominated by the long-lived dissipatons that are strongly coupled to the system.

%%%Xiao: non-Markovianity: long-lived dissipatons due to low-temperature or non-uniform energy structure of environment

Figure~\ref{fig1} illustrates the fermionic DQME theory, where the original OQS is mapped to a dissipaton-embedded system described in terms of the reduced density tensor (RDT), $\bm{\rho} = \rho(\vec{n},\vec{n}';\vec{m}^-,\vec{m}^+)$. Here, vectors $\vec{n}$ and $\vec{n}'$ describe system fermion configurations, while $\vec{m}^-$ and $\vec{m}^+$ represent electron-type and hole-type dissipaton configurations, respectively. The RDT elements with ($\vec{m}^-,\vec{m}^+) = \vec{0}$ yield the system's reduced density operator (RDO). As shown in Fig.~\ref{fig1}, each level can accommodate at most one fermion or dissipaton in accordance with the Pauli exclusion principle. 
In practice, a maximal dissipaton occupation number $M_{\rm max}$ is introduce to facilitate the numerical solution of the DQME \cite{li2024toward}.

%%%Xiao: use atomic unit? 
The fermionic DQME is given by (with $e = \hbar = 1$) \cite{li2024toward}:
\begin{align}
        \dot{\bm{\rho}} =\mathcal{L}\bm\rho=& -i[H_{_{\rm S}},\bm{\rho}] - \sum_j \left(
        \gamma_j^- \hat{N}_j \bm{\rho} + \gamma_j^+ \bm{\rho} \hat{N}_j\right) \notag \\
       & -i \sum_j \Big[ \big(\hat{c}_{\nu}^{\dagger} \hat{b}_{j} \bm{\rho} - \hat{b}_{j} \bm{\rho} \hat{c}_{\nu}^{\dagger} \big) + 
       \big(\hat{c}_{\nu} \bm{\rho} \hat{b}^{\dagger}_j -  \bm{\rho} \hat{b}_{j}^{\dagger} \hat{c}_{\nu} \big) \Big] \notag \\
       & -i \sum_j  \left[-\eta_{j}^{-}\, \hat{c}_{\nu} \hat{b}_j^{\dagger} \, \bm{\rho} - (\eta_{j}^{+})^\ast \,\hat{b}_j^{\dagger}\bm{\rho} \hat{c}_{\nu}\right]  \notag \\
       & -i \sum_j \left[ 
       \eta_{j}^{+} \, \hat{c}_{\nu}^{\dagger}\bm{\rho}\, \hat{b}_j
       + (\eta_{j}^{-})^\ast  \bm{\rho}\, \hat{b}_j \hat{c}_{\nu}^{\dagger}
       \right]. \label{eqn:DQME-SQ-1}
\end{align}
Here, $\hat{c}_{\nu}^{\dagger}$ ($\hat{c}_{\nu}$) creates (annihilates) a fermion at the $\nu$-th system fermion level, $\hat{b}_j^{\dagger}$ ($\hat{b}_j$) creates (annihilates) a dissipaton at $j$-th level, and $\hat{N}_j = \hat{b}_j^{\dagger} \hat{b}_j$ gives the dissipaton occupation. 
The RDO of the system, $\rho_0$, is obtained by projecting $\bm\rho$ onto the dissipaton vacuum. 

Fundamentally, dissipatons share similarities with previously proposed pseudomodes \cite{PhysRevA.55.2290,Tam18030402,Lam193721,cirio2023pseudofermion,Lin251289,PhysRevResearch.2.043058}, and the DQME is formally equivalent to the exact HEOM theory \cite{Jin08234703,PhysRevLett.109.266403} for reduced system dynamics.

\begin{figure}[t]
	\includegraphics[width=\columnwidth]{figure2.pdf}
	\caption{Structure of the neural network representing $\rho_{\rm pre} (\vec{n},\vec{n}';\vec{m})$. Boxes group nodes of the same type. The visible layer comprises $2N_{_{\rm S}} + N_{_{\rm E}}$ nodes, representing elements of $\vec{n}$, $\vec{n}'$, and $\vec{m}$, respectively. Connecting lines between boxes indicate that every node in one box is linked to all nodes in the other box through weighted connections. } \label{fig2}
\end{figure}

{\bf NQS representation of RDT.}
We now establish a NQS framework to characterize quantum correlations in OQSs. This framework begins by constructing a preliminary RDT as 
\begin{equation}
	\rho_{\rm pre} (\vec{n},\vec{n}';\vec{m}) = \sum_{\{\vec{a}\}}\psi(\vec{n};\vec{m};\vec{a}) \, \varphi^\ast (\vec{n}';\vec{m};\vec{a}),  \label{eqn:rho-pre-1}
\end{equation}
where $\vec{m} \equiv (\vec{m}^-,\vec{m}^+)$.
The wavefunctions $\psi$ and $\varphi$ are expressed using restricted Boltzmann machines (RBMs) \cite{PhysRevB.99.214306,PhysRevLett.122.250501,PhysRevLett.122.250502,PhysRevLett.122.250503, doi:10.1126/science.aag2302}, 
with $\vec{a}$ denoting $N_{\rm a}$ auxiliary nodes of binary values. This construction faithfully captures the mapping illustrated in Fig.~\ref{fig1}: the dissipaton states $\vec{m}$ encode the non-Markovian memory of the environment, while $\vec{a}$ accounts for the Markovian background responsible for the finite lifetimes of the dissipatons. Furthermore, this architecture inherently helps preserve the hermiticity and positive definiteness of the resulting RDO.

Specifically, 
\begin{equation}
\psi(\vec{n};\vec{m};\vec{a})=\sum_{\{\vec{h}\}} {\rm exp} [-E_{\vec{h}}(\vec{n};\vec{m};\vec{a})],
\end{equation}
where $\vec{h}$ represents $N_{\rm h}$ hidden nodes, and
\begin{align}
	& E_{\vec{h}}(\vec{n};\vec{m};\vec{a}) = 
	\bm{a}^T\bm{b}+\bm{n}^T\bm{c} +\bm{m}^T\bm{d}+\bm{h}^T\bm{g} +\bm{n}^T\bm{D}\bm{m} \nonumber \\
	& \qquad \quad +\bm{m}^T\bm{K}\bm{a}+\bm{n}^T\bm{X}\bm{h} +\bm{n}^T\bm{X}'\bm{a}+\bm{m}^T\bm{Y}\bm{h}.
 \label{eqn:Eh-1}
\end{align}
Here, $(\bm{b},\bm{c},\bm{d},\bm{g})$ denote complex-valued bias vectors, while $(\bm{D},\bm{K},\bm{X},\bm{X}',\bm{Y})$ represent complex-valued weight matrices of the RBM. An analogous RBM structure describes $\varphi(\vec{n}';\vec{m};\vec{a})$.

Figure~\ref{fig2} depicts the neural network architecture for
$\rho_{\rm pre}(\vec{n},\vec{n}';\vec{m})$. The complete analytic expression is provided in the Supplemental Material (SM) \cite{SM}. 
The dissipaton-related nodes in the visible layer explicitly encode the environmental memory content, in distinct contrast to conventional RBMs used to represent OQSs coupled to memoryless environments \cite{PhysRevLett.122.250501,PhysRevLett.122.250502,PhysRevLett.122.250503}.

The dimensionality of RDT can be reduced by leveraging its inherent symmetry and sparsity \cite{10.1063/1.4914514}. First, we enforce the symmetry constraint by constructing
\begin{equation}
    \rho_{\rm sym}(\vec{s}) = \rho_{\rm pre}(\vec{s})
    + (-1)^{\lfloor M^-/2 \rfloor+ \lfloor M^+/2 \rfloor} \rho_{\rm pre}^\ast(\vec{s}^{\, T}),
    \label{eqn:def-rhosym-1}
\end{equation}
where $\lfloor \cdot \rfloor$ denotes the floor function, $M^{\sigma} = \sum_j m_j^{\sigma}$, and $m_j^{\sigma}$ is the $j$-th element of $\vec{m}^\sigma$. Here, $\vec{s} \equiv (\vec{n},\vec{n}';\vec{m}^-,\vec{m}^+)$ represents a visible state, while $\vec{s}^{\, T} \equiv (\vec{n}',\vec{n};\vec{m}^+,\vec{m}^-)$ is its block-swapped counterpart.
Next, we introduce a filter function $f_{\rm{spa}}(\vec{s})$ that identifies, {\it a priori}, the zero elements of $\bm\rho$ based on the sparsity pattern dictated by the forms of $\Hs$ and $\Hse$ \cite{Hou15104112}.
In practice, $f_{\rm{spa}}(\vec{s})$ normally reduces the $\vec{s}$-space by over an order of magnitude \cite{SM}. The resulting symmetrized and filtered RDT is given by $\rho(\vec{s}) = f_{\rm spa}(\vec{s}) \rho_{\rm sym} (\vec{s})$.

Within the RBM representation, the RDT is expressed parametrically as $\bm\rho_{\bm\alpha}$, where $\bm\alpha$ denotes the complete set of RBM parameters. The total number of parameters is approximately $N_{\rm para} \approx 2N_{_{\rm S}}(N_{\rm h} + N_{\rm a}) + N_{_{\rm E}}(4N_{\rm h} + N_{\rm a})$. Provided that $N_{\rm h}$ and $N_{\rm a}$ scale linearly with $2N_{_{\rm S}} + N_{_{\rm E}}$, $N_{\rm para}$ scales approximately quadratically with the dimension of the dissipaton-embedded system \cite{SM}. This scaling is vastly more favorable than that of the original DQME, where the RDT dimension grows as $N_{\rm RDT} \approx 4^{N_{_{\rm S}}}  N_{_{\rm E}}^{M_{\rm max}}$.

{\bf Time evolution of NQS-DQME.}
%TDVP
Within the NQS framework, the exact DQME of Eq.~\eqref{eqn:DQME-SQ-1} is projected onto the parameter space.
Through the time-dependent variational principle \cite{PhysRevLett.122.250502,reh2021time}, the problem is transformed into determining the time evolution of the parameters $\bm\alpha(t)$. Numerically, this is achieved by minimizing the 2-norm $\Delta{s}^2= \Vert 
\dot {\bm{\rho}}_{\bm\alpha} - \mL {\bm{\rho}}_{\bm\alpha} \Vert^2_2$, which yields a linear equation for
$\dot{\bm\alpha}$ at each time step:
\begin{equation}
    \mathcal{S} \dot{\bm\alpha} = \mathcal{F}. 
    \label{eqn:tdvp-1}
\end{equation}
Here, $\mathcal{S}$ and $\mathcal{F}$ are a matrix and a vector defined in the parameter space, respectively. Evaluating their elements requires summming over all visible states $\vec{s}$, which we perform using the Markov chain Monte Carlo algorithm \cite{geyer1992practical}. The sampling process employs the Metropolis-Hastings criterion \cite{metropolis1953equation} with a customized probability distribution $p(\vec{s})\propto e^{-3 (M^+ + M^-)}$. The 
derivatives $\dot{\bm\alpha}$ are then obtained by solving \Eq{eqn:tdvp-1} \cite{minres}, and $\bm\alpha(t)$ are updated using the fourth-order Runge-Kutta algorithm \cite{press2007numerical, caoge2024code}.  
Further details are provided in the SM \cite{SM}.

To demonstrate the practicality and accuracy of the NQS-DQME approach, we investigate non-Markovian dynamics in two strongly correlated OQSs. We employ the numerically exact fermionic HEOM method implemented in the HEOM-QUICK2 program \cite{zhang2024heom} as the benchmark to provide reference values.

\begin{figure}[t]
	\includegraphics[width=\columnwidth]{figure3.pdf}
	\caption{
    (a) Schematic of open quantum dynamics for an impurity coupled to left ($L$) and right ($R$) reservoirs with chemical potentials $\mu_L$ and $\mu_R$. Gray shaded regions in the reservoirs represent Kondo clouds screening the impurity's localized spin.
    (b) $I_R(t)$ at different temperatures. The NQS-DQME results (solid lines, vertically offset for clarity) are benchmarked against HEOM reference values. Inset: steady-state current versus inverse temperature. 
    (c) Time evolution of $\Delta s^2$ for $k_{\rm B}T=0.3\,\Gamma$, computed with different $N_{\rm{h}} = N_{\rm{a}}$. Inset: $\mathcal{E}_{I_R}$ versus $N_{\rm{h}}$ with a power function fit.  
    (d) $I_R(t)$ at $k_{\rm B}T=0.3\,\Gamma$ computed with different $N_{\rm MC}$. Inset: $\mathcal{E}_{I_R}$ versus $N_{\rm MC}$.
    (e) Number of dynamical variables explicitly accessed in the NQS-DQME and HEOM methods versus $N_{_{\rm E}}$, where $N_{_{\rm E}}$ increases monotonically as $T$ decreases. Dashed lines are power function fits to the scattered data with $M_{\rm max} = 3$.    
    System parameters (in units of $\Gamma$): $\epsilon_0 = U_0/2 = 2$, $\Delta\epsilon = -7$, and $\Delta U = 6$; see the SM for further details \cite{SM}.
    }  \label{fig3}
\end{figure}

{\bf Case~1: non-Markovian charge relaxation with emergent Kondo correlations.}
We consider an OQS comprising a localized impurity symmetrically coupled to two noninteracting electron reservoirs. This setup is important for understanding electron transport through molecular junctions. The OQS is described by the single-impurity Anderson model \cite{And6141}. The impurity Hamiltonian is
$
H_{_{\rm S}}(t) = \epsilon_{0} (\hat{n}_\uparrow + \hat{n}_\downarrow)  + U_0\hat{n}_{\uparrow}\hat{n}_{\downarrow} + 
{\Theta}(t-t_0) [\Delta \epsilon (\hat{n}_\uparrow + \hat{n}_\downarrow) + \Delta U\hat{n}_{\uparrow}\hat{n}_{\downarrow}]
$,
where $\hat{n}_s$ is the occupation operator for spin-$s$ electrons, $\epsilon_0$ is the impurity energy, and $U_0$ is the Coulomb interaction energy.
At time $t_0 = 0$, a sudden quench shifts the impurity's parameters by $\Delta \epsilon$ and $\Delta U$, and a bias voltage is applied, establishing a chemical potential difference between the reservoirs. 
Figure~\ref{fig3}(a) depicts the ensuing  open quantum dynamics. The impurity level shift triggers electron transfer from the reservoirs to the impurity on a relatively short timescale. Subsequently, reservoir electrons redistribute their spins to screen the impurity's localized spin, leading to the formation of Kondo states at the impurity-reservoir interfaces over a longer timescale \cite{anders2005real,PhysRevLett.119.156601,Din24174120}. We denote $t_{\rm sh}$ and $t_{\rm lo}$ as representative times for the short-time and long-time regions, respectively.

Figure~\ref{fig3}(b) shows the time-dependent current $I_R(t)$ at various temperatures. At low temperatures,
$I_R(t)$ exhibits pronounced oscillations near $t = 1/\Gamma$, where $\Gamma$ is the impurity-reservoir hybridization strength. These transient oscillations are a characteristic signature of non-Markovian memory, arising from coherent impurity-reservoir coupling under suppressed
thermal fluctuations. The same coupling enhances Kondo correlations, leading to the increase in the steady-state current (evaluated at $t_{\rm lo}$) as temperature decreases, as seen in the inset.

Within the NQS-DQME framework, the expectation value of an observable $\hat{X}$, $X = \la \hat{X} \ra$, is evaluated from $\bm\rho_{\bm \alpha}$ \cite{SM}. The accuracy of the results is quantified by the relative error:
\begin{equation}
\mathcal{E}_X \equiv \frac{\int_{t_0}^{t_{{\rm lo}}} \lvert X(t) - X^{{{\rm ref}}}(t) \rvert \, {\rm d}t} { \int_{t_0}^{t_{{\rm lo}}} \frac{1}{2} \big(  \lvert X(t) \rvert + X^{{{\rm ref}}}(t) \rvert  \big) {\rm d} t } .
\end{equation}
As shown in Fig.~\ref{fig3}(b), the NQS-DQME results exhibit excellent agreement with the reference data, with $\mathcal{E}_{I_R} < 1\%$ across the entire temperature range. 

\begin{figure}[t]
	\includegraphics[width=\columnwidth]{figure4.pdf}
	\caption{Visualization of the RBM parameters for the open quantum dynamics shown in Fig.~\ref{fig3}. (a,b) Heatmaps of $\lvert \bm K(t_{\rm lo}) - \bm K(t_{\rm sh})\rvert$ at (a) $k_{\rm B}T = 3.0\,\Gamma$ and (b) $k_{\rm B}T = 0.3\,\Gamma$, for $t_{{\rm sh}} = 1.25/\Gamma$ and $t_{{\rm lo}} = 5/\Gamma$. The vertical axis indexes dissipaton levels (with decay rates sorted from slowest to fastest), and the horizontal axis indexes the auxiliary nodes. 
    (c,d) Time evolution of the auxiliary-layer-averaged weight difference for the $j$th dissipaton level, $\Delta K_j = \lvert K_j(t) - K_j(t_{\rm sh})\rvert$, at (c) $k_{\rm B}T = 3.0\,\Gamma$ and (d) $k_{\rm B}T = 0.3\,\Gamma$. Here, $\gamma_j \equiv {\rm Re}(\gamma^\sigma_j)$ denotes the decay rate in unit of $1/\Gamma$. The red line tracks the slowest-decaying dissipaton level.}
    %Identical $N_{\rm E}$ is adopted to enable a direct comparison between different temperatures. 
    \label{fig4}
\end{figure}

To explore the tradeoff between numerical cost and accuracy, we examine the low temperature regime ($k_{\rm B}T=0.3\,\Gamma$), where non-Markovian effects are prominent. The dissipaton-embedded system comprises $2$ system levels coupled inhomogeneously to $64$ dissipaton levels. Figure~\ref{fig3}(c) shows the evolution of the precision metric $\Delta s^2$ for different numbers of hidden nodes $N_{\rm h}$. The magnitude of $\Delta s^2$ decreases by orders of magnitude with a linear increase in $N_{\rm h}$, demonstrating rapid convergence. Similarly, Fig.~\ref{fig3}(d) shows that that accurate results are achieved with relatively few Monte Carlo samples ($N_{\rm MC}$).
Furthermore, Fig.~\ref{fig3}(e) 
demonstrates that the NQS-DQME approach  achieves high accuracy ($\Delta s^2 < 10^{-8}$) while utilizing drastically fewer dynamical variables than the conventional HEOM method. This comparison is conducted under equivalent conditions, as the same symmetry and sparsity relations are incorporated into both methods. These results collectively highlight the outstanding expressive power of the RBM for characterizing non-Markovian open quantum dynamics.

An important hypothesis in open quantum dynamics is that non-Markovian memory is crucial for long-time phenomena like the formation of Kondo-correlated states. However, directly verifying this by tracking dynamical variables in conventional methods such as the HEOM or the original DQME is prohibitively difficult due to their high dimensionality. In contrast, the NQS-DQME framework makes such tracking feasible, since the RDT is represented by weight matrices that form a much lower-dimensional parameter space.

This capability is demonstrated in Fig.~\ref{fig4}, which visualizes the evolution of the RBM weight matrix $\bm K$ connecting dissipaton-level nodes to auxiliary nodes for the open quantum dynamics shown in Fig.~\ref{fig3}. A comparison of the heatmaps in Fig.~\ref{fig4}(a) and (b) shows that the weight changes, $\lvert \bm K(t_{\rm lo}) - \bm K(t_{\rm sh})\rvert$, are significantly more pronounced for the slowest-decaying dissipaton level at the lower temperature. This is indicated by the darker top row in Fig.~\ref{fig4}(b). This observation is further confirmed by the time evolution of the auxiliary-layer-averaged weight difference $\Delta K_j(t)$ in Fig.~\ref{fig4}(c) and (d), where the slowest-decaying dissipaton level (red line) becomes increasingly prominent as temperature decreases, directly linking the non-Markovian memory to the emerging Kondo effect.

\begin{figure}
	\includegraphics[width=\columnwidth]{figure5.pdf}
	\caption{ (a) Schematic of quantum dissipative dynamics for two impurities coupled to a reservoir. Spin-exchange interaction (wavy line) and Kondo clouds (gray regions) are shown, where cloud extents indicate Kondo correlation strengths.
    Time evolution of (b) $\dot{S}_{12}$ and (c) $E_{\rm hyb}$ at various temperatures. 
    NQS-DQME results (solid lines, vertically offset for clarity) are benchmarked against HEOM reference values. 
    Inset in (c): $\Delta E_{\rm hyb} = E_{\rm hyb}(t_{\rm lo}) -E_{\rm hyb}(t_0)$ versus inverse temperature. The increase in $\Delta E_{\rm hyb}$ with decreasing temperature indicates weakened Kondo correlations in the long-time limit. System parameters (in units of $\Gamma$): $\epsilon_{0} = -U_0/2 =-6$, and $J=8$; see the SM for details \cite{SM}.
    } \label{fig5}
\end{figure}

%%%Xiao: more recent science papers on ac voltage control of coherent spin dynamics?
{\bf Case~2: non-Markovian spin relaxation with intertwined Kondo correlations.} 
We now examine a more complex OQS comprising two impurities coupled to a reservoir. This setup directly models recent experimental studies of coherent spin dynamics in surface molecules \cite{doi:10.1126/science.aay6779,doi:10.1126/science.abg8223}.
The system Hamiltonian is 
$
H_{_{\rm S}}(t) = \sum_{\nu=1,2} [\epsilon_0 (\hat{n}_{\nu \uparrow} +\hat{n}_{\nu\downarrow})
 + U_0 \hat{n}_{\nu\uparrow}\hat{n}_{\nu\downarrow}] 
 - \Theta(t-t_0) J \hat{\bm{S}}_1 \cdot \hat{\bm{S}}_2,
$
where the final term activates a ferromagnetic exchange interaction ($J > 0$) between the impurities at time $t_0$.

Figure~\ref{fig5}(a) shows the relaxation dynamics following the sudden activation of the exchange interaction. The initially independent localized spins on the two impurities gradually evolve towards parallel alignment. This results in a localized high-spin state that cannot be fully screened by the reservoir electrons. This underscreening progressively suppresses Kondo correlations during the temporal evolution \cite{PhysRevLett.103.197202,parks2010mechanical,li2020molecular}.

Figure~\ref{fig5}(b) and (c) depict the time derivative of the inter-impurity spin correlation, $\dot{S}_{12} = \frac{\partial }{\partial t}\la \hat{\bm S}_1 \cdot \hat{\bm S}_2\ra$, and the impurity-reservoir hybridization energy, $E_{\rm hyb} = \la H_{_{\rm SE}} \ra$, respectively, at various temperatures. We plot $\dot{S}_{12}$ rather than ${S}_{12}$ to highlight its more pronounced oscillatory behavior. At low temperatures, both quantities exhibit strong oscillations, a hallmark of non-Markovian memory effects.

The NQS-DQME results are in excellent agreement with the reference values, with maximum errors $\mathcal{E}_{{S}_{12}}$ and $\mathcal{E}_{{E}_{\rm hyb}}$ both below $1\%$ for $t_{\rm lo} > 3/\Gamma$. This quantitative agreement is remarkable, as the spin relaxation dynamics are governed by the complex interplay of direct inter-impurity correlations and impurity-reservoir Kondo correlations. Furthermore, similar to Case~1, the NQS representation achieves this accuracy with a significantly reduced number of dynamical variables compared to conventional approaches \cite{SM}.

{\bf Summary.} We have developed a NQS framework for the efficient simulation of non-Markovian open quantum dynamics by integrating the dissipaton concept into the ansatz. This NQS-DQME approach provides a compact and physically interpretable epresentation of the RDT, enabling the simultaneous treatment of complex environmental memory and many-body correlations with significantly enhanced computational efficiency.

While the current framework shares a common challenge with many NQS methods, i.e., the computationally demanding minimization of the loss function $\Delta s^2$ \cite{PhysRevLett.122.250503}.
This bottleneck can be addressed using advanced optimizers and more sophisticated network architectures. Given its proven scalability and interpretability, the NQS-DQME approach presents a promising and versatile platform for exploring rich quantum phenomena in correlated, non-Markovian settings.

\begin{acknowledgments}

Support from the National Natural Science Foundation of China (Grants No.\ 22393912, No.\ 22321003, No.\ 22103073, No.\ 22173088, and No.\ 22373091), the Innovation Program for Quantum Science and Technology (Grants No.\ 2021ZD0303301 and No.\ 2021ZD0303306), the AI for Science Foundation of Fudan University (Grant No. FudanX24AI023), and the Strategic Priority Research Program of Chinese Academy of Sciences (Grant No.\ XDB0450101) is gratefully acknowledged. 

\end{acknowledgments}

\bibliographystyle{aiptit}
\bibliography{bibrefs}
% \bibliographystyle{aiptit}
% \bibliographystyle{aip}
% \bibliography{bibrefs}

% \begin{thebibliography}{10}

% \end{thebibliography}

\end{document}

\title{Simulating Non-Markovian Open Quantum Dynamics with Neural Quantum States}
\author{Long~Cao} \altaffiliation{These authors contributed equally to this work.}
\affiliation{Hefei National Research Center for Physical Sciences at the Microscale, 
University of Science and Technology of China, Hefei, Anhui 230026, China}

\author{Liwei~Ge} \altaffiliation{These authors contributed equally to this work.}
\affiliation{Hefei National Research Center for Physical Sciences at the Microscale, 
University of Science and Technology of China, Hefei, Anhui 230026, China}

\author{Daochi~Zhang}
\affiliation{Department of Chemistry, Fudan University, Shanghai 200438, China}

\author{Xiang~Li}
\affiliation{Hefei National Research Center for Physical Sciences at the Microscale, 
University of Science and Technology of China, Hefei, Anhui 230026, China}

\author{Jialin Pan}
\affiliation{Hefei National Research Center for Physical Sciences at the Microscale, 
University of Science and Technology of China, Hefei, Anhui 230026, China}

\author{Yao~Wang}
\affiliation{Hefei National Research Center for Physical Sciences at the Microscale, 
University of Science and Technology of China, Hefei, Anhui 230026, China}

\author{Rui-Xue~Xu}
\affiliation{Hefei National Research Center for Physical Sciences at the Microscale, 
University of Science and Technology of China, Hefei, Anhui 230026, China}
\affiliation{Hefei National Laboratory, Hefei, Anhui 230088, China}

\author{YiJing~Yan}
\affiliation{Hefei National Research Center for Physical Sciences at the Microscale, 
University of Science and Technology of China, Hefei, Anhui 230026, China}

\author{Xiao~Zheng} \email{xzheng@fudan.edu.cn}
\affiliation{Department of Chemistry, Fudan University, Shanghai 200438, China}
\affiliation{Hefei National Laboratory, Hefei, Anhui 230088, China}

\date{\today}

\noindent {\bf Supplemental Material for}

\maketitle

%%%%%%%%%%%%%%%%%%%

%%%%%%%%%%%%%%%%%%%%%
%%%%%%%%%%%%%%%%%%%
\renewcommand{\thesection}{S\arabic{section}}
\renewcommand{\thesubsection}{S\arabic{section}.\arabic{subsection}}
\renewcommand{\thesubsubsection}{S\arabic{subsection}.\arabic{subsubsection}}
\renewcommand{\theequation}{S\arabic{equation}}
\renewcommand{\thetable}{S\arabic{table}}
\renewcommand{\thefigure}{S\arabic{figure}}
\renewcommand{\thepage}{S\arabic{page}}
%%%%%%%%%%%%%%%%%%%%%

\tableofcontents
\clearpage

% \section{Fermionic dissipaton-embedded quantum master equation in second quantization method}
\section{Quantum dynamical equations for many-body open quantum systems}

\subsection{Open quantum system coupled to fermionic environment}

We consider an open quantum system (OQS) in which the system is linearly coupled to the environment consisting of several noninteracting fermionic particle reservoirs. 
The Hamiltonian of the total system is described by (with $e = \hbar  = 1$)
\begin{align}
\Ht & = \Hs + \He + \Hse, \\
\He & = \sum_\alpha \sum_k  \epsilon_{\alpha k} \hat{d}^{\dagger}_{\alpha k}\hat{d}_{\alpha k}, \\
\Hse & = \sum_{\nu=1}^{N_{_{\rm S}}} \sum_\alpha 
	\left(\hat{c}_{\nu}^{\dagger}\hat{F}_{\alpha \nu} + \hat{F}_{\alpha \nu}^{\dagger}\hat{c}_{\nu}\right), 
\end{align}
where $\Hs$ represents the system of primary interest, $\He$ denotes the environment, and $\Hse$ describes the system-environment coupling.
In these terms, $\hat{c}^\dag_{\nu}$ ($\hat{c}_{\nu}$) denotes the creation (annihilation) operator for the $\nu$-th single-particle state of the system,  $\hat{d}^\dag_{\alpha k}$ ($\hat{d}_{\alpha k}$) denotes the creation (annihilation) operator for the $k$-th single-particle state of the $\alpha$-th reservoir, 
where the indices $\nu$  and $k$ include the spin degree of freedom.
$\hat{F}_{\alpha \nu}=\sum_k t_{\nu \alpha k}\hat{d}_{\alpha k}$, where $\hat{d}^\dag_{\alpha k}$ ($\hat{d}_{\alpha k}$) represents the particle creation (annihilation) operator for the $k$-th single-particle state of $\alpha$-reservoir, and $t_{\nu \alpha k}$ denotes the coupling strength between the system's $\nu$-th state and the $\alpha$-reservoir's $k$-th state.

\subsection{Characterizing the non-Markovian memory of environment}
The hybridization spectral functions of the $\alpha$-reservoir are defined as
\begin{equation}
    J_{\alpha \nu }(\omega) \equiv \pi \sum_k \left\lvert t_{\nu \alpha k} \right\rvert^2
    \delta(\omega-\epsilon_{\alpha k}).
\end{equation}
In this work, we adopt a Lorentzian form for the hybridization spectral functions, i.e.,
\begin{equation}
    J_{\alpha \nu}(\omega) = \frac{\Gamma_{\nu\alpha} W_\alpha^2 }{(\omega-\Omega_{\alpha})^2+W_\alpha^2}.  
\end{equation}
Here, $\Gamma_{\nu\alpha}$ is the hybridization strength between the system's $\nu$-th state and the $\alpha$-reservoir, $\Gamma_\nu = \sum_\alpha \Gamma_{\nu\alpha}$, and $\Omega_\alpha$ and $W_\alpha$ are the band center and width of the $\alpha$-reservoir, respectively. 
For simplicity, we take $\Omega_\alpha = \mu_\alpha$, where $\mu_\alpha$ is the chemical potential of the $\alpha$-reservoir. In particular, we set $\mu_\alpha^{\rm eq} = 0$ in thermal equilibrium.

Since a  reservoir of  noninteracting fermionic particles linearly coupled to the system follows Gaussian statistics, its influence on the reduced system dynamics is entirely captured by the reservoir hybridization correlation functions, which are defined as
\begin{align}
    C_{\alpha \nu}^{+}(t-\tau) & \equiv \langle 
        \hat{F}_{\alpha \nu}^{\dagger}(t) \hat{F}_{\alpha \nu}(\tau) \rangle_{_{\rm E}}, \nonumber \\ 
    C_{\alpha \nu}^{-}(t-\tau) & \equiv \langle
        \hat{F}_{\alpha \nu}(t) \hat{F}^{\dagger}_{\alpha \nu}(\tau) \rangle_{_{\rm E}},
\end{align}
where $\hat{F}_{\alpha \nu}(t) = e^{i \He t}\hat{F}_{\alpha \nu}e^{-i \He t}$, and $\la \cdot \ra_{_{\rm E}} = {\rm tr}( \cdot \rho^{\rm eq}_{_{\rm E}})$ with $\rho^{\rm eq}_{_{\rm E}}$ being the density matrix of the decoupled environment in thermal equilibrium. 
The reservoir correlation functions are related to the hybridization spectral functions through the fluctuation-dissipation theorem
\begin{equation}
    C_{\alpha \nu}^{\sigma}(t) = 
    \int_{-\infty}^{\infty} {\rm d} \omega \, e^{\sigma i \omega t} 
    f^{\sigma}_{\alpha}(\omega) J_{\alpha \nu}(\omega), \label{eqn:FD}
\end{equation}
where $f^{\sigma}_{\alpha}(\omega) = 1/(1+e^{\sigma \beta_{\alpha} (\omega-\mu_{\alpha})})$, and $\beta_{\alpha} = 1/(k_{\rm B}T)$ is the inverse temperature of the $\alpha$-reservoir. 
By employing a sum-over-poles expansion for $f^{\sigma}_{\alpha}(\omega)$ and $J_{\alpha \nu}(\omega)$, the reservoir correlation functions can be expressed as a linear combination of exponential functions
\begin{equation}
	C_{\alpha \nu}^{\sigma}(t) = \sum_{p=1}^{P} 
    \eta_{\alpha \nu p}^{\sigma} e^{-\gamma_{\alpha \nu p}^{\sigma}t}.
    \label{eqn:Ct-exp-1}
\end{equation}
The $j$ index used in the main text corresponds to the multi-index $\{\alpha \nu p\}$ here. For the benchmark tests presented in the main text, the \Pade spectral decomposition scheme \cite{hu2010communication} is employed to carry out the exponential expansion in \Eq{eqn:Ct-exp-1}.

When a time-dependent voltage is applied to the $\alpha$-reservoir, the energy levels and chemical potential of the reservoir experiences a homogeneous shift $\Delta \mu_{\alpha}(t)$, which introduces a phase factor to the non-stationary reservoir correlation functions:
\begin{equation}
    \tilde{C}_{\alpha \nu}^{\sigma}(t,\tau) = 
    {\rm exp} \left[ \sigma i \int_{\tau}^t {\rm d} t' \, \Delta\mu_{\alpha}(t') \right] 
    C_{\alpha \nu}^{\sigma}(t-\tau).  \label{eqn:Ct-factor-1}
\end{equation}
The two-time hybridization correlation functions $\tilde{C}_{\alpha \nu}^{\sigma}(t,\tau)$ serve as the memory kernel that explicitly express the non-Markovian memory content of the $\alpha$-reservoir. Equations~\eqref{eqn:Ct-exp-1} and \eqref{eqn:Ct-factor-1} provide the foundation for several numerically exact theories, such as the hierarchical equations of motion (HEOM) method and the dissipaton-embedded quantum master equation (DQME) in a second quantized form, for characterizing non-Markovian dissipative dynamics of many-body OQSs.

\subsection{Detailed forms of fermionic HEOM and DQME}

%We introduce the DQME-SQ method by relating it to the numerically exact fermionic HEOM method \cite{Jin08234703,PhysRevLett.109.266403}.  

%The HEOM method provides a numerically exact description of the non-Markovian dissipative dynamics of many-body OQS. 

The basic variables of HEOM are the auxiliary density operators (ADOs) \cite{Jin08234703}. In the path-integral formulation, ADOs are defined by acting Grassmannian generating fields $\{\mB^\sigma_j\}$ on the initial-time reduced density operator (RDO) $\rho_0$, as follows \cite{li2024toward}
\begin{align}
\rho^{(-\ldots-+\ldots+)}_{p_1\ldots p_I q_1 \ldots q_J} & \equiv (-1)^{(I+J)} 
\int_{t_0}^t \mD {\bm x} \mD {\bm x'} \, e^{i S_f[\bm x]}\, e^{-i S_b[\bm x']} \nonumber \\
& \quad \times e^{-\Phi[\bm x, \bm x']}\, \mB^-_{p_1} \cdots \mB^-_{p_I} \mB^+_{q_1} \cdots \mB^+_{q_J} \rho_0.  \label{ado-1}
\end{align}
Here, $S_f$ and $S_b$ denote forward and backward action functionals, respectively, $\Phi$ is the Feynman-Vernon influence functional, and $\mB^-_j$ ($\mB^+_j$) represents the transfer of a fermionic particle from the system (environment) to the environment (system) via the $j$-th dissipation level, where $j$ is an abbreviation for the multi-index $\{\alpha \nu p\}$.
Because of the Grassmann anti-symmetry inherent to fermions, interchanging any two $\mB^\sigma_j$ fields in \Eq{ado-1} yields a minus sign.

With the exponential decomposition of reservoir memory in \Eq{eqn:Ct-exp-1}, the HEOM are expressed as
\begin{align}
	\dot{\rho}^{(-\ldots-+\ldots+)}_{p_1\ldots p_I q_1 \ldots q_J} &=
	\bigg[ -i\mL_{_{\rm S}}+  \sum_{j}\gamma_j
	\bigg] \rho^{(-\ldots-+\ldots+)}_{p_1\ldots p_I q_1 \ldots q_J} \nonumber \\ 
    & \quad +  \sum_{r=1}^{I} \mC^{-}_{p_r}
	\rho^{(-\ldots--\ldots+)}_{p_1\ldots p_{r-1}  p_{r+1} \ldots q_J}
	+\sum_{r=1}^{J} \mC^{+}_{q_r}
	\rho^{(-\ldots++\ldots+)}_{p_1\ldots q_{r-1}  q_{r+1} \ldots q_J} \nonumber \\
     & \quad +  \sum_j \mA^{-}_j
	\rho^{(-\ldots--+\ldots+)}_{p_1\ldots p_I j q_1 \ldots q_J} 
	+\sum_j \mA^{+}_j 
	\rho^{(-\ldots-++\ldots+)}_{p_1\ldots p_I j q_1 \ldots q_J}. 
 \label{eqn:heom-1}
\end{align}
Here, $\mL_{_{\rm S}} = [H_{_{\rm S}},\cdot ]$ denotes the system  Liouvillian, and $\mA_j^{\sigma}$ and $\mC^{\sigma}_{p}$ are tier-up and tier-down superoperators acting on the ADOs, respectively.
% see Ref.~\onlinecite{Han18234108} for detailed derivations. 

The DQME theory provides a fresh viewpoint on the ADOs. The entire set of ADOs is mapped onto the reduced density tensor (RDT) of the dissipaton-embedded system, denoted as $\bm{\rho}$. Specifically, the mapping from an ADO to an element of the RDT is given by 
\begin{equation}
  \rho^{(-\ldots-+\ldots+)}_{p_1\ldots p_I q_1 \ldots q_J} \mapsto
  \rho(\vec{n},\vec{n}';\vec{m}^{-},\vec{m}^{+}), \label{N1}
\end{equation}
where each of the vectors $\vec{n}$ and $\vec{n}'$ corresponds to a fermionic particle configuration for the system, while $\vec{m}^\sigma$ represents a dissipaton configuration. 

%\clearpage
The formal equivalence between the HEOM and DQME is established by writing the equation of motion for each element of the RDT, as follows,
%The detailed form of DQME is given by
%\begin{equation}
\begin{align}
    & \dot\rho(\vec{n},\vec{n}';\vec{m}^-,\vec{m}^+) = 
       -i \sum_{\vec{k}} \left[ H_{_{\rm S}}(\vec{n},\vec{k})
        \rho(\vec{k},\vec{n}',\vec{m}^-,\vec{m}^+)-
        \rho(\vec{n},\vec{k},\vec{m}^-,\vec{m}^+)
        H_{_{\rm S}}(\vec{k},\vec{n}') 
          \right] 
    \nonumber \\
    & +\sum_j(\gamma^-_j m^-_j+\gamma^+_j m^+_j)  \rho(\vec{n},\vec{n}';\vec{m}^-,\vec{m}^+) 
    \nonumber \\
    & -i\sum_j (-1)^{\sum_{k>j}m^-_k}
        \sum_{\vec{k}} \left[c_\nu^\ast(\vec{k},\vec{n})
    	\rho(\vec{k},\vec{n}',\vec{m}^-_{j+},\vec{m}^+)-(-1)^{M^-+M^+}
    \rho(\vec{n},\vec{k},\vec{m}^-_{j+},\vec{m}^+)
    	c_\nu^\ast(\vec{n}',\vec{k}) \right]
    \nonumber \\
    & -i\sum_j (-1)^{\sum_{k<j}m^+_k} \sum_{\vec{k}}
        \left[(-1)^{M^-+M^+} c_\nu(\vec{n},\vec{k})
            \rho(\vec{k},\vec{n}',\vec{m}^-,\vec{m}^+_{j+})-
            \rho(\vec{n},\vec{k},\vec{m}^-,\vec{m}^+_{j+})
            c_\nu(\vec{k},\vec{n}') \right] 
    \nonumber \\
    & -i\sum_j (-1)^{\sum_{k<j}m^-_k}  \sum_{\vec{k}}
    \Big[ (-1)^{M^--1} \, \eta_{j}^{(-)} c_\nu(\vec{n},\vec{k})
        \rho(\vec{k},\vec{n}',\vec{m}^-_{j-},\vec{m}^+)
    \nonumber \\
    & \qquad    -(-1)^{M^+} \big[\eta_{j}^{(+)} \big]^* 
        \rho(\vec{n},\vec{k},\vec{m}^-_{j-},\vec{m}^+)
        c_\nu(\vec{k},\vec{n}')
      \Big]
    \nonumber \\
    & -i\sum_j (-1)^{\sum_{k>j}m^+_k} \sum_{\vec{k}}
    \Big[    (-1)^{M^-}\eta_{j}^{(+)} c_\nu^\ast(\vec{k},\vec{n})
        \rho(\vec{k},\vec{n}',\vec{m}^-,\vec{m}^+_{j-})
    \nonumber \\
    & \qquad  -(-1)^{M^+-1}
       \big[ \eta_{j}^{(-)} \big]^*   \rho(\vec{n},\vec{k},\vec{m}^-,\vec{m}^+_{j-})
        c_\nu^\ast(\vec{n}',\vec{k})   \Big] 
        \label{eqn:dqme-sq}
\end{align}
where $M^\sigma=\sum_k m^\sigma_k$, $c_\nu(\vec{n},\vec{n}')$ denotes the matrix representation of the annihilation operator $\hat{c}_\nu$, with $\nu$ belonging to the multi-index $j =\{\alpha \nu p\}$, and $\vec{m}_{j+(-)}^{\sigma}$ indicates that adding (subtracting) a dissipaton to (from) the $j$-th level in the dissipaton configuration $\vec{m}^{\sigma}$.

% \section{Restricted Boltzmann machine representation of reduced density tensor}
\section{Neural quantum state representation of RDT}

In the restricted Boltzmann machine (RBM) presentation, the auxiliary wavefunctions are expressed as
\begin{align}  
  \psi(\vec{n};\vec{m};\vec{a}) & = \sum_{\{\vec{h}\}}
	{\rm exp}  \Big( \bm{a}^T\bm{b}+\bm{h}^T\bm{g}+\bm{n}^T\bm{c}+ \bm{m}^T\bm{d} 
        \nonumber \\ 
        & \qquad  \qquad \quad  + \bm{n}^T\bm{X}\bm{h}+ \bm{n}^T\bm{X}'\bm{a}+\bm{m}^T\bm{Y}\bm{h}+
     \bm{m}^T\bm{K}\bm{a}+\bm{n}^T\bm{D}\bm{m} \Big)  \nonumber \\ 
        & = {\rm exp} \Big(\bm{a}^T\bm{b}+\bm{n}^T\bm{c}+\bm{m}^T\bm{d}+ \bm{n}^T\bm{X}'\bm{a}
         \bm{m}^T\bm{K}\bm{a}+ \bm{n}^T\bm{D}\bm{m} \Big)  \notag \\
        & \quad \times \prod_j \bigg[  1+{\rm exp} \Big(g_j+\sum_in_iX_{ij}+\sum_im_iY_{ij} \Big)
         \bigg],  \label{eqn:psi-1}\\
    % \notag \\
	\varphi(\vec{n}';\vec{m};\vec{a}) & = \sum_{\{\vec{h}\}} {\rm exp}
     \Big(\bm{a}^T\bm{b}+\bm{h}^T\bm{g}+\bm{n}'^T\bm{c}+ \bm{m}^T\bm{d}   \notag \\
        & \qquad \qquad \quad + \bm{n}'^T\bm{X}\bm{h}+\bm{n}'^T\bm{X}'\bm{a}+\bm{m}^T\bm{Y}'\bm{h}+
      \bm{m}^T\bm{K}\bm{a}+\bm{n}'^T\bm{D}\bm{m} \Big)  \notag \\
        & \quad = {\rm exp}  \Big(\bm{a}^T\bm{b}+\bm{n}'^T\bm{c}+\bm{m}^T\bm{d}+
      \bm{n}'^T\bm{X}'\bm{a}+\bm{m}^T\bm{K}\bm{a}+ \bm{n}'^T\bm{D}\bm{m}\Big)  \notag \\
        & \quad \times \prod_j \bigg[  1+{\rm exp} \Big(g_j+\sum_i n'_i X_{ij}+\sum_i m_i Y'_{ij} \Big)
          \bigg].   \label{eqn:phi-1}
\end{align}
Here, each of the vectors $\vec{n}$ and $\vec{n}'$ corresponds to fermionic particle configuration of the system, while $\vec{m}$ represents a dissipaton configuration; $\vec{a}$ and $\vec{h}$ denote a set of auxiliary nodes and hidden nodes, respectively; $(\bm{b},\bm{c},\bm{d},\bm{g})$ and $(\bm{D},\bm{K},\bm{X},\bm{X}',\bm{Y},\bm{Y}')$ 
represent the bias vectors and weight matrices of RBMs, respectively. 
%The $\bm{c}$ here is different from the annihilation operators in the DQME equation.
%
Then, the neural quantum state (NQS) representation for the RDT of the DQME theory is established as 
\begin{align}
	\rho_{{\rm pre}} (\vec{n},\vec{n}';\vec{m}) 
        & = 
	\sum_{ \{\vec{a} \}}\psi(\vec{n};\vec{m};\vec{a})
	\varphi^* (\vec{n}';\vec{m};\vec{a}) \notag \\
	&= {\rm exp}\Big[ \sum_i(d_i+d^*_i)m_i+\sum_i(c_in_i+c^*_in'_i)+ \bm{n}^T\bm{D}\bm{m}+\bm{n}'^T\bm{D}'^*\bm{m}
        \Big] \notag \\
	&\quad \times \prod_j \bigg[      1+{\rm exp} 
        \Big(b_j+b^*_j+\sum_in_iX'_{ij}+   \sum_in'_iX'^*_{ij}+    \sum_i m_i \big(K_{ij}+K^*_{ij} \big)\Big)
	\bigg] 	\notag \\
	&\quad \times \prod_j \bigg[ 
         1+{\rm exp}
         \Big(g_j+\sum_in_iX_{ij}+\sum_im_iY_{ij} \Big)
         \bigg]       \notag \\
	&\quad \times \prod_j \bigg[ 
         1+{\rm exp}
         \Big(g^*_j+\sum_in'_iX^*_{ij}+  \sum_im_iY'^*_{ij} \Big)
         \bigg]. \label{eqn:rhopre-1}
\end{align}
% Constructing RDT from auxiliary wavefunctions can guarantee the hermiticity and positive semidefiniteness of 
%
In Eqs.~\eqref{eqn:psi-1}-\eqref{eqn:rhopre-1}, $\psi$ and $\varphi$ share the same weights associated with the visible nodes representing the system particle configurations (the $\vec{n}$ nodes). This preserves the Hermiticity and positivity of the RDO.

The RDT is subject to the following symmetry constraint:
\begin{equation}
 \rho(\vec{s})
    = (-1)^{\lfloor M^-/2 \rfloor+ \lfloor M^+/2 \rfloor} \rho^\ast(\vec{s}^{\, T}),
    \label{eqn:def-rhosym-1}
\end{equation}
where $\lfloor \cdot \rfloor$ denotes the floor function, $M^{\sigma} = \sum_j m_j^{\sigma}$, and $m_j^{\sigma}$ is the $j$-th element of $\vec{m}^\sigma$. Here, $\vec{s} \equiv (\vec{n},\vec{n}';\vec{m}^-,\vec{m}^+)$ represents a visible state, while $\vec{s}^{\, T} \equiv (\vec{n}',\vec{n};\vec{m}^+,\vec{m}^-)$ is its block-swapped counterpart.
We enforce the symmetry constraint by constructing
\begin{equation}
    \rho_{\rm sym}(\vec{s}) = \rho_{\rm pre}(\vec{s})
    + (-1)^{\lfloor M^-/2 \rfloor+ \lfloor M^+/2 \rfloor} \rho_{\rm pre}^\ast(\vec{s}^{\, T}).
    \label{eqn:def-rhosym-1}
\end{equation}
The number of real-valued parameters in the RBM is
\begin{equation}
	N_{{\rm para}} = 2N_{{\rm h}}+
	N_{{\rm a}}+2(1+N_{{\rm h}}+N_{{\rm a}})N_{_{\rm S}}+
	(1+4N_{{\rm h}}+N_{{\rm a}})N_{_{\rm E}} + 4N_{_{\rm E}} N_{_{\rm S}}.
\end{equation}
%

%the coefficients before $N_{_{\rm S}}$ and $N_{_{\rm E}}$ are different so as  to satisfy the Hermiticity and positivity of the RDO. Figure~2 in the main text exhibits a bilateral symmetry in the neural network parameters except the four weights associated with the visible nodes representing $\vec{m}$.

\section{Exploiting the sparsity of RDT} \label{sec:sparsity}

Harnessing the sparsity of the RDT significantly enhances the numerical efficiency of the NQS-DQME approach. This is because a substantial portion of the RDT elements, represented by the visible states ${\vec{s}}$ of the RBM, are inherently zero. Identifying and excluding these zero-valued elements during sum-over-states operations substantially reduces the computational cost of NQS-DQME. 

The sparsity pattern of the RDT is determined by the forms of $\Hs$ and $\Hse$. To establish this pattern, we categorize all visible states $\vec{s}$ based on the characteristic vector $\bm u(\vec{s})$, defined as $u_\nu = (M_{\nu}^+ - M_{\nu}^-) + (n_{\nu} - n_{\nu}')$, where $\nu=1,\ldots,N_{_{\rm S}}$, and $M_{\nu}^{\sigma} = \sum_{\alpha p}m^{\sigma}_{\alpha \nu p}$. Visible states associated with the same $\bm u$-vector belong to the same category. In cases where a category includes an RDO element (i.e., $\rho(\vec{n}, \vec{n}'; \vec{m} = \vec{0})$), then $\rho(\vec{s}) = 0$ for all $\vec{s}$ in that category, provided the matrix element $\la \vec{n}' \vert \exp(-\Hs) \vert \vec{n} \ra = 0$. Conversely, if a category contains no RDO element, then $\rho(\vec{s}) = 0$ for all $\vec{s}$ in that category.

The establishment of the sparsity pattern enables the implementation of an efficient filter, denoted as $f_{_{\rm spa}}(\vec{s})$, which allows for rapid screening of visible states before processing in the NQS-DQME solver. This filter promptly returns a value of $0$ if $\rho(\vec{s})$ is determined to be zero-valued based on the aforementioned rules.

The efficiency of the sparsity filter is demonstrated in Table~\ref{tab:sparsity}. The filter reduces the number of non-zero elements in the RDT by approximately an order of magnitude across a range of temperatures, leading to a significant reduction in computational cost.

\begin{table}[h]
\centering
\caption{Reduction of dynamical variables via the sparsity filter for the single-impurity system (Case 1). The number of non-zero RDT elements, $N_{\rm RDT}$, is shown with and without the application of the sparsity filter for different temperatures.}  \label{tab:sparsity}
\begin{tabular*}{\textwidth}{@{\extracolsep\fill}lccccc}
\hline\hline
$k_{\rm B}T/\Gamma$  & $3.0$  & $2.0$  & $1.0$  & $0.6$  & $0.3$  \\
\hline
$N_{\rm RDT}$ (full) & 52880 & 99536 & 167760 & 261648 & 385296 \\
$N_{\rm RDT}$ (filtered) & 6724 & 12884 & 21988 & 34612 & 51332 \\
Reduction Factor & 7.9 & 7.7 & 7.6 & 7.6 & 7.5 \\
\hline\hline
\end{tabular*}
\end{table}

\section{Temporal evolution of neural network parameters}  \label{sec:tdvp}

The DQME can be formally recast into a compact form of
\begin{equation}
    \dot{\bm{\rho}} = \mL \bm{\rho}, \label{eqn:heom-2}
\end{equation}
where $\mL$ represents a generalized Liouvillian of the dissipaton-embedded system. In the NQS representation, the temporal evolution of $\bm\rho$ is transformed into the evolution of the time-dependent parameters of the neural network $\{\bm\alpha(t)\}$, i.e.,
\begin{equation}
    \dot{\bm{\rho}}  \simeq \sum_{k} \dot{\alpha}_k\, \frac{\partial \bm\rho}{\partial {\alpha_k}}. \label{eqn:dot-rho-1}
\end{equation}
Here, $\frac{\partial \bm\rho}{\partial {\alpha_k}}$ are obtained through analytic differentiation, and the values of $\dot{\alpha}_k$ are determined by applying the time-dependent variational principle (TDVP) \cite{PhysRevLett.122.250502}, 
which minimizes the 2-norm loss function:
\begin{equation}
    \Delta{s}^2= \Big\Vert \sum_k \dot{\alpha}_k \hat{O}^k {\bm{\rho}} - \mL {\bm{\rho}} \Big\Vert^2_2,    
\end{equation}
where $\hat{O}^k \equiv \frac{\partial}{\partial {\alpha_k}}$. 
Minimizing $\Delta{s}^2$ involves solving the linear equation
\begin{equation}
    \sum_k \mathcal{S}_{l,k} \, \dot{\alpha}_k = \mathcal{F}_l, \label{SF}
\end{equation}
where
\begin{align}
    \mathcal{S}_{l,k} & =\overline{O^\ast_lO_k}+ \overline{O^\ast_kO_l} = 2\, {\rm Re} \left(\overline{O^\ast_lO_k} \right), \\ 
    \mathcal{F}_l & =\overline{O_l^\ast {\mathcal{L}}_{\rm loc}}+\overline{{\mathcal{L}}^\ast_{\rm loc} O_l} = 2\, {\rm Re} \left(\overline{O_l^\ast {\mathcal{L}}_{\rm loc}} \right),
\end{align}
with the definitions:
\begin{align}
    \overline{{A} {B}} & \equiv\sum_{ \{\vec{s}\} } {A}(\vec{s}) {B}(\vec{s}),  \\
    {O}_k(\vec{s}) & \equiv\frac{\partial \rho(\vec{s})}{\partial{\alpha_k}}, \\
	{\mathcal{L}}_{\rm loc}(\vec{s}) & \equiv \langle\vec{s} |\mL| \bm\rho\rangle. 
\end{align}
At each time step, Eq.~\eqref{SF} is solved either directly by regularizing
the matrix $\mathcal{S}$ (adding a small positive constant $\delta \sim 10^{-12}- 10^{-8}$ to its diagonal elements to eliminate its singularity), or iteratively using the minimum residual quadratic linear problem (MINRES-QLP) algorithm \cite{minres}. The resulting time derivatives $\{\dot{\alpha}_k\}$ are then used to update $\{\alpha_k\}$ via the fourth-order Runge-Kutta (RK4) algorithm. 
However, solving Eq.~\eqref{SF} with high accuracy at each time step is computationally demanding for complex systems, representing a major bottleneck in the current NQS-DQME implementation.

%{\color{blue}In practice, we introduce a small perturbation $\delta*I$ to the calculated $\mathcal{S}$ to eliminate the sigularity of $\mathcal{S}$ with $\delta \approx 10^{-8}\  to\ 10^{-12}$ for different cases.}

%{\color{gray}the minimum residual quadratic linear problem (MINRES-QLP) algorithm \cite{minres}}
%

%%%Xiao: reference for the MCMC sampling? 
The implementation of TDVP typically requires a Markov chain Monte Carlo (MCMC) sampling algorithm \cite{geyer1992practical} to perform the sum-over-states operations. In this work, inspired by the dissipaton occupation number truncation in the fermionic DQME method, we limit the sampling to a truncated visible-state space with a maximum allowed occupancy of dissipatons set to $M_{\rm max}$. Consequently, any RDT elements $\rho(\vec{s})$ with $M(\vec{s}) = \sum_{\sigma j}m^\sigma_j > M_{\rm max}$ are enforced to be zero and automatically excluded from the MCMC sampling. Additionally, the sparsity filter $f_{\rm spa}(\vec{s})$ introduced in \Sec{sec:sparsity} further accelerates the sampling procedure. 
%to facilitate a direct comparison with the fermionic HEOM method

Based on extensive applications of the HEOM method, it has been found that ADOs at lower tiers hold relatively greater significance and less amount compared to those at higher tiers. In the context of DQME, this suggests that the sampling procedure should prioritize the RDT elements associated with lower dissipaton occupancy. Based on this notion, we design a MCMC sampling protocol that is compatible with the NQS-DQME framework, which proceeds as follows.

The set of accessible states $Q = \{\vec{s}\mid M(\vec{s}) \leqslant M_{\rm max}\}$ partitions into two subsets, the traversed states $Q_1$ and the sampling sampling $Q_2$, where $Q_1 = \{\vec{s}\mid M(\vec{s}) \leqslant L\}$  with $L < M_{\rm max}$ and $Q_2= Q\setminus Q_1$. The sum-over-states operation is then carried out as
\begin{equation}
    \sum_{ \{\vec{s}\} } A(\vec{s})  \simeq \sum_{\vec{s} \in Q} A(\vec{s})  
     = \sum_{\vec{s} \in Q_1} A(\vec{s}) + \sum_{\vec{s} \in Q_2} A(\vec{s}).
     \label{eqn:sum-Q}
\end{equation}
Here, summing over the $Q_1$ subset is straightforward and does not require any sampling due to its small size. For the sum over the $Q_2$ subset, the following equation is used:
\begin{equation}
     \sum_{\vec{s}\in Q_2} {A}(\vec{s}) = \sum_{\vec{s}\in Q_2} \tilde{p}(\vec{s}) \, \frac{{A}(\vec{s})}{\tilde{p}(\vec{s})}  = Z_p \left\langle  \frac{A(\vec{s})}{\tilde{p}(\vec{s})}  \right\rangle_{Q_2}.
 \label{eqn:sum-Q2-1}
\end{equation}
Unlike conventional quantum Monte Carlo methods, where the distribution function $\tilde{p}(\vec{s})$ is typically determined by the normalized wavefunction or density matrix, here the unnormalized RDT employs $\tilde{p}(\vec{s}) = e^{-\lambda M(\vec{s})}$ for sampling in the $Q_2$ space, with the empirical parameter $\lambda = 3$. The average of $X(\vec{s})$ over the $Q_2$ subset is computed as $\la X(\vec{s}) \ra_{Q_2} = \sum_{\vec{s}\in Q_2} X(\vec{s}) p(\vec{s})$, where $p(\vec{s})=\tilde{p}(\vec{s})/Z_p$, with $Z_p = \sum_{\vec{s}\in Q_2} \tilde{p}(\vec{s})$ being the partition function.  
To navigate through the $Q_2$ space efficiently, the MCMC walker proposes a new state $\vec{s}'$ from the current state $\vec{s}$ by flipping a pair of elements ($0$ to $1$ or $1$ to $0$ interchangeably) using one of two randomly chosen rules: (i) simultaneously flip both $m_j^+$ and $m_{j}^-$ for a given $j$, or (ii) simultaneously flip $m_j^+$ (or $m_j^-$) and its corresponding $n_{\nu}$ (or $n_{\nu}'$), where $\nu$ belongs to the multi-index $j$. 

%Unlike conventional quantum Monte Carlo methods, where the distribution function $p(\vec{s})$ is typically determined by the normalized wavefunction or density matrix, here we enforce $p(\vec{s})=e^{-\lambda M(\vec{s})}$ for sampling in the $Q_2$ space, with the empirical parameter $\lambda = 3$.
The Metropolis-Hastings criterion \cite{metropolis1953equation} is then applied to accept or reject the proposed move, ensuring convergence to the target distribution.

To illustrate the source of the high efficiency of this sampling method, we compute the variance of the sampled quantity $\sum_{\vec{s} \in Q} A(\vec{s})$. First, we define the random variables $A_{2,i}$ and $A_2$:
\begin{align}
    P\left(A_{2,i}=\frac{A(\vec{s})}{\tilde{p}(\vec{s})}\right) = p(\vec{s}),\\
    A_2 = \frac{1}{N_{\rm MC}}\sum_{i=1}^{N_{\rm MC}} A_{2,i}.
\end{align}
Here, $A_{2,i}$ denotes the $N_{\rm MC}$ independent identically distributed random variables obtained from MC sampling. The value we compute via MC sampling is actually the random variable $A$:
\begin{align}
    A  = 
    A_1 +Z_p * A_2,
\end{align}
where $A_1 = \sum_{\vec{s} \in Q_1} A(\vec{s})$.
It is easy to see that:
\begin{equation}
    \sum_{\vec{s} \in Q} A(\vec{s}) = \operatorname{E}[A],
\end{equation}
where $\operatorname{E}[A]$ denotes the expectation value of $A$. The efficiency of this sampling can be characterized by the variance of the random variable $A$:
\begin{equation}
    \operatorname{Var}(A) = Z_p^2 * \frac{\operatorname{Var} (A_{2,i})}{N_{\rm MC}}.
\end{equation}
It can be seen that as the sampling truncation $L$ for $Q_2$ increases, $Z_p$ decreases; moreover, since the value of $A(\vec{s})$ often significantly decreases with increasing occupation number of dissipatons, $\operatorname{Var} (A_{2,i})$ also decreases as $L$ increases. In contrast, conventional full-space MC sampling corresponds to a truncation $L=-1$. Therefore, for the same number of samples, the variance of this sampling method is smaller than that of conventional full-space MC sampling, indicating that it has higher sampling efficiency.
The Table~\ref{tab:variace} shows the variance of the sampled quantity $\Vert\mathcal{L}\bm{\rho}\Vert^2_{2}$ at the beginning of the temporal evolution against the sampling truncation $L$ in the Case~1 at the $k_{\rm B}T=0.3\,\Gamma$, where $A(\vec{s}) = \left|(\mathcal{L}\bm{\rho})(\vec{s})\right|^2$. This table demonstrates that the our sampling strategy with the sampling truncation $L=2$ can achieve the same variance as the full space MC sampling with a sampling number four orders of magnitude smaller, rendering the remarkable smooth evolution curve in Fig.~3(b). This strategy can be extended to spin-lattice systems and quantum chemistry by combining a direct summation of the mean‑field results and low‑excitation configurations with Monte Carlo sampling of the high‑excitation configurations.

\begin{table}[ht]
\centering
\caption{The variance of the sampled $\Vert\mathcal{L}\bm{\rho}\Vert^2_{2}$ at $t=0./\Gamma$ in Case 1 at the $k_{\rm B}T=0.3$.}  \label{tab:variace}
\begin{tabular*}{\columnwidth}{@{\extracolsep\fill}lcccc}
\hline\hline
$L$   &-1   &  0    &1  &2   \\
\hline
        $Z_p$   &23.93  &    19.93    &13.55  &5.94\\
        $\operatorname{Var} (A_{2,i})$  &$3.09*10^{-3}$    
        &  $3.68*10^{-3}$   
        &$1.9313*10^{-4}$  &$8.98*10^{-6}$\\
        $Z_p^2\operatorname{Var} (A_{2,i})$ &   $1.76$    & $1.46$ 
        &$3.55*10^{-2}$       & $3.17*10^{-4}$\\
\hline\hline
\end{tabular*}
\end{table}

When the open quantum dynamics begins from an equilibrium state of the full system, the initial parameters $\bm\alpha(t_0)$ are determined by minimizing the $\Vert\mathcal{L} \bm{\rho}_{\bm \alpha}\Vert^2_2$ or  by evolving a random initial state to the long-time limit.

\section{Evaluation of physical quantities}

With the RDT obtained using the NQS-DQME solver, a wide range of physical observables of experimental relevance can be evaluated. 
In the following, we explicitly include the spin index $s$. Thus, the index $\nu$ used in the preceding sections and from the main text now extends to $\nu s$, and the multi-index becomes $j = \{\alpha \nu s p\}$.
For instance, the occupation number of spin-$s$ electrons on the $\nu$-th impurity ($n_{\nu s}$),  the inter-impurity spin correlation ($S_{12}$), and the von Neumann entropy ($S_{{\rm vN}}$) are calculated based on the RDO,
$\rho_0(\vec{n}, \vec{n}') = \rho(\vec{n}, \vec{n}'; \vec{m} = \vec{0})$, as follows:
\begin{align}
	n_{\nu s} &= {\rm tr}_{_{\rm S}} \left( \hat{n}_{\nu s} \rho_0 	\right), \nonumber \\
	S_{12} &= {\rm tr}_{_{\rm S}} \big( \hat{\bm{S}}_1 \cdot \hat{\bm{S}}_2 \rho_0\big),  \nonumber \\
	S_{{\rm vN}} &= -{\rm tr}_{_{\rm S}} \left(\rho_0 {\rm log} \rho_0\right),
\end{align} 
where $\hat{n}_{\nu s} = \hat{c}^{\dagger}_{\nu s} \hat{c}_{\nu s}$,
and $\hat{\bm{S}}_{\nu} = \frac{1}{2} \sum_{ss'} \hat{c}^{\dagger}_{\nu s} \bm{\sigma}_{ss'} \hat{c}_{\nu s'}$ with $\bm{\sigma}=(\sigma_x,\sigma_y,\sigma_z)$ the Pauli matrices.

The electric current flowing into $\alpha$-reservoir ($I_\alpha$) and impurity-reservoir hybridization energy ($E_{\rm hyb}$) are calculated using the RDT elements with $M(\vec{s}) = 1$ as follows:
\begin{align}
     I_{\alpha} &= -2\, {\rm Im} \bigg\{
    \sum_{\nu s p} \sum_{\vec{n},\vec{n}'}   
    \la \vec{n}' \vert \hat{c}_{\nu s} \vert \vec{n} \ra \,
     \rho(\vec{n},\vec{n}';\vec{m}^{+}_{\alpha \nu s p},\vec{0}) \bigg\}, \\
     E_{\rm hyb} &= 2\,  {\rm Re} \bigg\{
    \sum_{\alpha \nu s p}  \sum_{\vec{n},\vec{n}'}
     \la \vec{n}' \vert \hat{c}_{\nu s} \vert \vec{n} \ra \,
     \rho(\vec{n},\vec{n}';\vec{m}^{+}_{\alpha \nu s p},\vec{0}) \bigg\},
\end{align} 
where $\vec{m}^{+}_{\alpha \nu s p}$ denotes the vector $\vec{m}^+$ with only one nonzero element, i.e., ${m}^{+}_{\alpha \nu s p} = 1$.

\section{Details of numerical calculations}

\subsection{Energetic parameters of OQSs and RBM hyperparameters}

In Case~1, the values of energetic parameters adopted are (in units of $\Gamma=(\Gamma_{L}+\Gamma_{R})/2$):
$\Gamma_{L} = \Gamma_{R} = 1$, $\epsilon_0 = U_0/2 = 2$, $\Delta\epsilon = -7$, $\Delta U = 6$, $W_{L} = W_R = 50$, $\Delta\mu_{L} = 0$, and $\Delta\mu_{R} = 0.5$. 

In Case~2, the values of energetic parameters adopted are (in units of $\Gamma=\Gamma_1/2$):
$\Gamma_1 = \Gamma_2 = 2$, $\epsilon_{0} = - U_0/2 = -6$, $J = 8$, and $W = 20$.

The number of \Pade poles employed to decompose the hybridization correlation functions, as well as the setting of RBM hyperparameters, for different temperatures in Cases 1 and 2 are given in Table~\ref{pole1} and Table~\ref{pole2}, respectively. 

\begin{table}[h]
\centering
\caption{Number of \Pade poles and RBM hyperparameters employed in Case 1.}  \label{pole1}
\begin{tabular*}{\textwidth}{@{\extracolsep\fill}lccccc}
\hline\hline
$k_{\rm B}T/\Gamma$  & $3.0$  & $2.0$  & $1.0$  & $0.6$  & $0.3$  \\
\hline
$P_{\Pade}$  & 3  & 4  & 5  & 6  & 7  \\
$N_{\rm h}$  & 20 & 20 & 20 & 20 & 30 \\
$N_{\rm a}$  & 20 & 20 & 20 & 20 & 30 \\
$N_{\rm para}$&3712&4584&5456&7783&10510\\
\hline\hline
\end{tabular*}
\end{table}

\begin{table}[h]
\centering
\caption{Number of \Pade poles and RBM hyperparameters  employed in Case 2.} \label{pole2} 
\begin{tabular*}{\textwidth}{@{\extracolsep\fill}lccccc}
\hline\hline
$k_{\rm B}T/\Gamma$ & $2.40$  & $0.48$  & $0.40$  & $0.32$  & $0.24$  \\
\hline
$P_{\Pade}$  & 3  & 4  & 4  & 5  & 5  \\
$N_{\rm h}$  & 60 & 60 & 60 & 70 & 70 \\
$N_{\rm a}$  & 20 & 20 & 20 & 30 & 30 \\
$N_{\rm para}$ &9652&11868&11868&16674&16674\\
\hline\hline
\end{tabular*}
\end{table}

The dimensions of the dissipaton-embedded system (number of system and dissipaton levels) in Cases 1 and 2 are shown in Tables~\ref{size_case1} and \ref{size_case2}, respectively. The number of dissipaton levels $N_{_{\rm E}}$ increases as the temperature decreases, which is necessary to accurately capture the non-Markovian memory effects at lower temperatures.

\begin{table}[h]
\centering
\caption{Dimensions of the dissipaton-embedded system for Case 1.}  \label{size_case1}
\begin{tabular*}{\textwidth}{@{\extracolsep\fill}lccccc}
\hline\hline
$k_{\rm B}T/\Gamma$  & $3.0$  & $2.0$  & $1.0$  & $0.6$  & $0.3$  \\
\hline
$N_{_{\rm S}}$ & 2 & 2 & 2 & 2 & 2 \\
$N_{_{\rm E}}$  & 32 & 40 & 48 & 56 & 64 \\
\hline\hline
\end{tabular*}
\end{table}

\begin{table}[h]
\centering
\caption{Dimensions of the dissipaton-embedded system for Case 2.}  \label{size_case2}
\begin{tabular*}{\textwidth}{@{\extracolsep\fill}lccccc}
\hline\hline
$k_{\rm B}T/\Gamma$  & $2.40$  & $0.48$  & $0.40$  & $0.32$  & $0.24$  \\
\hline
$N_{_{\rm S}}$ & 4 & 4 & 4 & 4 & 4 \\
$N_{_{\rm E}}$ & 32 & 40 & 40 & 48 & 48 \\
\hline\hline
\end{tabular*}
\end{table}

\subsection{Monte Carlo sampling for the space of visible states}
%836+5888

%3024+48308
%Figure~{mc1} presents a comparison between the simulation results obtained using the MCMC sampling and those obtained by explicitly exploring the full space of visible states. The two sets of results agree excellently with each other. 

In our practical implementation, we employ a hybrid sampling strategy. The subspace of visible states $\{\vec{s}\}$ with $M(\vec{s}) \leqslant L$ (subspace $Q_1$) is fully enumerated, while the subspace with $L < M(\vec{s}) \leqslant M_{\rm max}$ (subspace $Q_2$) is sampled using the MCMC algorithm. For both Case 1 and Case 2, we set $M_{\rm max} = 3$ and $L = 2$, which ensures sufficient accuracy for the DQME calculations presented in this work.
The sizes of the traversed space $N_{Q_1}$ and the sampling space $N_{Q_2}$ for all temperatures studied are provided in Table~\ref{subset1} and Table~\ref{subset2}.

As shown in Fig.~3(d) of the main text, MCMC sampling with $N_{\rm MC} = 4096$ states achieves high accuracy ($\mathcal{E}_{I_R} < 1\%$) even at the lowest temperature studied. This sampling size constitutes only a small fraction of the $Q_2$ subspace dimension ($N_{Q_2} = 48128$), demonstrating the high efficiency of our customized MCMC algorithm.

% {\color{blue} The results in Fig.~3(b) and (c) in the main text were obtained with $N_{\rm MC} \geqslant 8000$ samples. For example, in Case~1, the total number of visible states is 5888 with the $L=3$ truncation at $k_{\rm B}T = 3.0\,\Gamma$. We sampled 4 Markov chains with each consisting of 2000 states. 
% For $k_{\rm B}T = 0.3\,\Gamma$, the total number of sampling states is 48128, and we sampled 5 Markov chains, each consisting of 4000 states. The size of traversed space $Q_1$ and the sampling space $Q_2$ of Case~1 is listed in Table~\ref{subset2}. In Case~2, the total number of visible states is 26624 with the $L=3$ truncation at $k_{\rm B}T = 2.4\,\Gamma$.  We sampled 4 Markov chains with each consisting of 2500 states.}

Figure~\ref{mc2} presents a comparison between the results obtained via MCMC sampling and the reference values from the HEOM method. Despite minor sampling errors,
the MCMC approach maintains small relative errors: $\mathcal{E}_{S_{12}}\approx0.7\%$ at $k_{\rm B}T=2.4\,\Gamma$ and $\mathcal{E}_{S_{12}}\approx1.5\%$ at $k_{\rm B}T=0.24\,\Gamma$. The inset shows that the relative error decreases rapidly with increasing  sample size $N_{\rm MC} $, achieving a relatively low error $\mathcal{E}_{S_{12}}\approx1.7\%$ with a small sampling number $N_{\rm MC}=12288$ compared to the full $Q_2$ space size $N_{Q_2}=91008$.

%The occurrence of the burr on the $\dot{S}_{12}(t)$ line is attributed to the derivation process applied to $S_{12}(t)$. In particular, as the number of samples increases, the accuracy of the MCMC results improves. 

% \clearpage

% \begin{figure}[h]
% 	\includegraphics[width=\textwidth]{mc_case1.pdf}
% 	\caption{Temporal evolution of (a) $I_R(t)$ and (b) $n_{s}(t)$ ($s=\uparrow$ or $\downarrow$) at two different temperatures in Case 1. The results obtained by using the MCMC algorithm are compared with those by explicitly exploring the full space of visible states. } \label{mc1}
% \end{figure}
%
%\clearpage

\begin{table}[h]
\centering
\caption{Size of the traversed space and the sampling space in Case 1.}  \label{subset1}
\begin{tabular*}{\textwidth}{@{\extracolsep\fill}lccccc}
\hline\hline
$k_{\rm B}T/\Gamma$  & $3.0$  & $2.0$  & $1.0$  & $0.6$  & $0.3$  \\
\hline
$N_{Q_1}$  & 836  & 1284  & 1828  & 2468  & 3204  \\
$N_{Q_2}$  & 5888 & 11600 & 20160 & 32144 & 48128 \\
\hline\hline
\end{tabular*}
\end{table}

\begin{table}[h]
\centering
\caption{Size of the traversed space and the sampling sapce in Case 2.}  \label{subset2}
\begin{tabular*}{\textwidth}{@{\extracolsep\fill}lccccc}
\hline\hline
$k_{\rm B}T/\Gamma$  & $2.40$  & $0.48$  & $0.40$  & $0.32$  & $0.24$  \\
\hline
$N_{Q_1}$  & 3842  & 5898  & 5898  & 8394  & 8394  \\
$N_{Q_2}$  & 26624 & 52400 & 52400 & 91008 & 91008 \\
\hline\hline
\end{tabular*}
\end{table}

\begin{figure}[h]
	\includegraphics[width=0.8\textwidth]{mc_case2.pdf}
     \caption{ Temporal evolution of $\dot{S}_{12}(t)$  at $k_{\rm B}T = 2.4\,\Gamma$ and $0.24\,\Gamma$ in Case 2. The lines are vertically offset for clarity. Inset:  variation of $\mathcal{E}_{{S}_{12}}$ with the sampling number $N_{\rm MC}$ at $k_{\rm B}T = 0.24\,\Gamma$. RBM hyperparameters: $N_{\rm h}=70$ and $N_{\rm a}=30$.}  \label{mc2}
\end{figure}

\clearpage

\subsection{Numerical accuracy of the NQS-DQME results}

%\Tab{error1} and \Tab{error2} present the relative errors of various quantities from the benchmark tests for Cases~1 and 2, respectively.  
%The relative error for a quantity $X$ is defined as
%\begin{equation}
%\mathcal{E}_X \equiv \frac{\int_{t_0}^{t_{{\rm lo}}} \lvert X(t) - X^{{{\rm ref}}}(t) \rvert \, {\rm d}t} { \int_{t_0}^{t_{{\rm lo}}} \left( \lvert  X^{{{\rm ref}}}(t) \rvert  + \lvert X(t) \rvert \right) {\rm d} t /2 } .
%\end{equation} 
%
%The reference value $X^{{{\rm ref}}}(t)$ is obtained from the fermionic HEOM method.

\Tab{error1} and \Tab{error2} present the relative errors of key quantities for Cases~1 and 2, respectively, benchmarking the NQS-DQME results against reference HEOM data. A clear trend is observed: the relative errors generally increase as the temperature decreases, where non-Markovian effects become more pronounced. This trend underscores the growing challenge of accurately capturing quantum correlations in many-body OQSs as the environment's memory content increases.

Particularly, in Case 1 at the lowest temperature ($k_BT=0.3\Gamma$), the relative error for $I_R(t)$ obtained via MCMC sampling (0.965\%) is slightly lower than that from full sampling (0.998\%). This minor discrepancy may be attributed to a partial cancellation between the errors introduced by the stochastic MCMC sampling and those inherent to the RBM representation itself.

\begin{table}[h]
\centering
\caption{Relative errors of the NQS-DQME results for $I_{R}(t)$ and $n_{\uparrow}(t)$ in Case~1 with $t_{\rm lo} = 3.8/\Gamma$. } \label{error1}
\begin{tabular*}{\textwidth}{@{\extracolsep\fill}llccccc}
\hline\hline
& \multicolumn{4}{r}{$k_{\rm B}T/\Gamma$} \\\cmidrule{3-7}
Error  & sampling  & $3.0$  & $2.0$  & $1.0$  & $0.6$  & $0.3$  \\
\hline
$\mathcal{E}_{I_{\rm R}}$ (\%)  & full  & 0.020  & 0.047  & 0.108  & 0.400  & 0.998  \\
$\mathcal{E}_{n_{\uparrow}}$ (\%)  & full  & 0.000  & 0.001  & 0.016  & 0.032  & 0.027  \\
$\mathcal{E}_{I_{\rm R}}$ (\%)  & MCMC  & 0.070  & -  & -  & -  & 0.965  \\
$\mathcal{E}_{n_{\uparrow}}$ (\%)  & MCMC  & 0.001  & -  & -  & -  & 0.047  \\
\hline\hline
\end{tabular*}
%\extrafootertext{Note: The integral error is defined as $E^{\rm int}_O \equiv \int_{t_0}^{t_{{\rm long}}} \lvert O^{{\rm calc}}(t) - O^{{{\rm ref}}}(t) \rvert {\rm d}t / \int_{t_0}^{t_{{\rm long}}} \lvert  O^{{{\rm ref}}}(t) \rvert {\rm d} t$, where $O$ is the evaluated quantity, $t_0 = 0$.}
\end{table}

\begin{table}[h]
\centering
\caption{Relative errors of the NQS-DQME results for $S_{\rm 12}(t)$, $S_{\rm vN}(t)$ and $E_{\rm hyb}(t)$ in Case~2 with $t_{\rm lo} = 3/\Gamma$. } \label{error2}
\begin{tabular*}{\textwidth}{@{\extracolsep\fill}llccccc}
\hline\hline
& \multicolumn{4}{r}{$k_{\rm B}T/\Gamma$} \\\cmidrule{3-7}
Error  & sampling  & $2.40$  & $0.48$  & $0.40$  & $0.32$  & $0.24$  \\
\hline
$\mathcal{E}_{S_{\rm 12}}$ (\%)  & full  & 0.305  & 0.631  & 0.770  & 0.902  & 0.828  \\
% $E^{\rm int}_{\dot{S}_{\rm 12}}$ (\%)  & full  & 1.389  & 3.643  & 3.844  & 4.923  & 6.996  \\
$\mathcal{E}_{S_{\rm vN}}$ (\%)  & full  & 0.125  & 0.262  & 0.202  & 0.311  & 0.267  \\
$\mathcal{E}_{E_{\rm hyb}}$ (\%)  & full  & 0.049  & 0.212  & 0.123 & 0.156  & 0.286  \\
$\mathcal{E}_{S_{\rm 12}}$ (\%)  & MCMC  & 0.746  & -  & -  & -  & 1.567  \\
% $E^{\rm int}_{\dot{S}_{\rm 12}}$ (\%)  & MCMC  & 3.17  & -  & -  & -  & -  \\
$\mathcal{E}_{S_{\rm vN}}$ (\%)  & MCMC  & 0.328  & -  & -  & -  & 0.612  \\
$\mathcal{E}_{E_{\rm hyb}}$ (\%)  & MCMC  & 0.088  & -  & -  & -  & 0.316  \\
\hline\hline
\end{tabular*}
\end{table}

Figure~\ref{fig:spectral} shows the spectral function $A(\omega) = A_{\nu s}(\omega) = \frac{1}{2\pi} \int dt e^{i\omega t} \langle \{\hat{c}_{\nu s}(t), \hat{c}^\dag_{\nu s}\}\rangle$ for a localized impurity in Case~2. Comparing initial ($t = t_0$) and final ($t = t_{\rm lo}$) states at temperatures $k_{\rm B}T = 2.4\,\Gamma$ and $0.24 \,\Gamma$, we observe: (i) At low temperature ($k_{\rm B}T = 0.24\,\Gamma$), the final state exhibits significant reduction of the Kondo peak height at $\omega = 0$. (ii) This reduction indicates that high-spin state formation in the two-impurity composite creates underscreened Kondo states with weakened impurity-reservoir correlations \cite{PhysRevLett.103.197202,parks2010mechanical,li2020molecular}.
(iii) At high temperature ($k_{\rm B}T = 2.4\,\Gamma$), thermal fluctuations completely suppress Kondo correlations.

\begin{figure}[t]
    \centering
	\includegraphics[width=0.65\textwidth]{Aw.pdf}
    \caption{Impurity spectral function in the initial and final states in Case 2.} \label{fig:spectral} 
\end{figure}

\begin{figure}[h]
    \centering
    \includegraphics[width=0.85\textwidth]{case2_SM.pdf}
    \caption{(a) Number of explicitly accessed dynamical variables by the NQS-DQME and HEOM methods at various temperatures in Case~2, where the values of $N_{_{\rm E}}$ correspond to the different temperatures; see \Tab{pole2} for details. (b) Temporal evolution of the von Neumann entropy $S_{\rm vN}$ at various temperatures in Case 2. The results obtained by the NQS-DQME are compared to the reference values from the HEOM method. The lines are vertically offset for clarity.} \label{fig:case2_SM}
\end{figure}

\clearpage

Figure~\ref{fig:case2_SM}(a) demonstrates that, similar to the trend shown in Fig.~3(e) of the main text, the NQS-DQME approach substantially reduces the number of dynamical variables required to describe non-Markovian dynamics compared to the HEOM method. Figure~\ref{fig:case2_SM}(b) depicts the temporal evolution of the von Neumann entropy $S_{\rm vN}(t)$ at various temperatures in Case 2.
It is evident that, regardless of the temperature, the final equilibrium state exhibits lower von Neumann entropy compared to the initial state, signifying correlated orientations of the localized spins on the two impurities. This observation is consistent with Fig.~3(e) in the main text.

% It is evident that, regardless of the temperature, the final equilibrium state exhibits higher hybridization energy compared to the initial state. This increase in hybridization energy signifies a weakened Kondo correlation, which is consistent with the observation in \Fig{fig:spectral}.

% \clearpage

\subsection{Visualizing the evolution of RBM parameters}

\begin{figure}[h]
    \centering
    \includegraphics[width=0.85\textwidth]{interpretability.pdf}
    \caption{Temporal evolution of auxiliary-layer-averaged weight per dissipaton
    level, $\Delta K_j$, at various temperatures: (a) $k_{\rm B}T=3.0\,\Gamma$, (b) $k_{\rm B}T=1.0\, \Gamma$, (c) $k_{\rm B}T=0.6\,\Gamma$, and (d) $k_{\rm B}T=0.3\,\Gamma$. Here, the index $j$ labels the decomposition index $p$ in Eq.~\eqref{eqn:Ct-exp-1}, i.e.,  $\Delta K_j(t) = \frac{1}{N}\sum_{\alpha \nu a}\lvert K_{\alpha \nu j a}(t) - K_{\alpha \nu j a}(t_{\rm sh})\rvert$ where the multi-index $\{\alpha \nu\}$ is the same as that in Eq.~\eqref{eqn:Ct-exp-1} and $a$ represents the auxiliary nodes. The $\Delta \bm K$ matrix  in the main text is averaged only over multi-index $\{\alpha \nu\}$.   \label{fig:interpretability}}
\end{figure}

Figure~\ref{fig:interpretability} supplements Fig.~4 in the main text with additional intermediate temperatures. As the temperature decreases, the slowest-decaying dissipaton level progressively dominates the long-time relaxation dynamics. This trend is clearly visible in \Fig{fig:interpretability}, where the red line representing this level gradually becomes increasingly prominent among all lines. 

\section{Comparing the performance of NQS-DQME to MPS-HEOM}

%%%Xiao: we preliminary comparison between rbm and tns, where mps is a kind of tns 

Matrix product states (MPSs), a class of tensor network states (TNSs), have been used to provide compact representations of the dynamical variables (the RDO and ADOs) in the HEOM method \cite{Shi18174102}. In the MPS-HEOM approach, an ADO is approximated as
\begin{equation}
    \rho_{\vec{m}}^{nn'} \approx \sum_{i_0,i_1,\cdots,i_{N_{_{\rm E}}+2}} B_{0}(i_0,n,i_1)B_{1}(i_1,m_1,i_2)\cdots B_{N_{_{\rm E}}}(i_{N_{_{\rm E}}},m_{N_{_{\rm E}}},i_{N_{_{\rm E}}+1})B_{N_{_{\rm E}}+1}(i_{N_{_{\rm E}}+1},n',i_{N_{_{\rm E}}+2}),
\end{equation}
where $n$ and $n'$ index system levels, $\vec{m}$ labels the principal dissipation modes, and $i_k$ (for $k=0, 1, \cdots, N_{_{\rm E}}+2$) are the auxiliary tensor indices. The ranks of these tensors (the range of $i_k$) are the bond dimensions $N_{\rm bond}$, which determine the number of dynamical variables in the MPS-HEOM method. Importantly, $N_{\rm bond}$ depends sensitively on both the strength of many-body correlations and the complexity of the non-Markovian memory in the OQS.

\begin{figure}[h]
	\includegraphics[width=\textwidth]{rbm_mps.pdf}
%	\caption{Comparison of the evolution of the occupation number of the up spin electron calculated by the MPS-HEOM method and the NQS-DQME method on Case 1 at different temperatures. Parameters: $N_{\rm a}=N_{\rm h}$ for RBM; the bond is the bond dimension (rank) of MPS. }
\caption{Comparative performance of the MPS-HEOM and NQS-DQME methods for the single-impurity system (Case 1). Time evolution of the spin-up occupation number $n_\uparrow(t)$ at (a) high temperature ($k_{\rm B}T=3.0\,\Gamma$) and (b) low temperature ($k_{\rm B}T=0.3\,\Gamma$). Results from the NQS-DQME method with different numbers of hidden nodes ($N_{\rm h}$) are compared with those from the MPS-HEOM method with different bond dimensions ($N_{\rm bond}$). The NQS-DQME method achieves higher accuracy with significantly fewer parameters than the MPS-HEOM approach (see Table~\ref{mps} for more details).
}
\label{fig:mps}
\end{figure}

\begin{table}[h]
\centering
\caption{Comparison of parameter counts and accuracy for the MPS-HEOM and NQS-DQME methods in Case 1.}  \label{mps}
\begin{tabular*}{\textwidth}{@{\extracolsep\fill}lccccc}
\hline\hline
$k_{\rm B}T/\Gamma$  & $3.0$  & $0.3$  \\
\hline
$N_{\rm para\,(RBM)}$  & 634 ($N_{\rm h}=N_{\rm a} =2$, $\mathcal{E}_{n_{\uparrow}}=0.07\%$)   & 5545 ($N_{\rm h}=N_{\rm a} = 15$, $\mathcal{E}_{n_{\uparrow}}=0.11\%$)  \\
$N_{\rm para\,(MPS)}$  & 9184 ($N_{\rm bond}=12$, $\mathcal{E}_{n_{\uparrow}}=0.81\%$)  & 299120 ($N_{\rm bond}=50$, $\mathcal{E}_{n_{\uparrow}}=0.97\%$) \\
$N_{\rm RDT}$  &6724   &51332 \\
\hline\hline
\end{tabular*}
\end{table}

We now perform a direct comparison between the MPS-HEOM method \cite{Shi18174102} and our NQS-DQME approach by evaluating their performance in simulating the non-Markovian dynamics of the single-impurity Anderson model (Case 1).

Figure~\ref{fig:mps} presents a comparative analysis of the time evolution for the spin-up occupation number, $n_\uparrow(t)$, calculated by both methods at different temperatures. Panel (a) shows the high-temperature regime  ($k_{\rm B}T = 3.0\,\Gamma$). 
The MPS-HEOM method requires a bond dimension of 12 (corresponding to 9,184 parameters) to achieve a truncation error of $\mathcal{E}_{n_{\uparrow}}=0.81\%$. In contrast, the RBM within the NQS-DQME framework achieves a significantly lower error of $0.07\%$ using only 634 parameters ($N_{\rm h}=N_{\rm a} =2$).

This efficiency gap becomes even more pronounced at low temperatures, as shown in panel (b) for $k_{\rm B}T = 0.3\,\Gamma$. Here, MPS-HEOM requires a substantially larger bond dimension of 50 (299,120 parameters) yet yields a higher error of $0.97\%$. The RBM approach, with 5,545 parameters ($N_{\rm h}=N_{\rm a} =15$), maintains superior accuracy with an error of only $0.11\%$.

The quantitative parameter counts and corresponding errors are summarized in Table~\ref{mps}. The data clearly demonstrates that the NQS-DQME method achieves higher accuracy with orders of magnitude fewer parameters than MPS-HEOM across both temperature regimes. This underscores the enhanced representational capacity and superior scalability of the RBM ansatz for capturing the complex non-Markovian dynamics of many-body OQSs.

%Figure~\ref{fig:mps} compares the results of the MPS-HEOM method and the NQS-DQME method in Case 1 with different numbers of parameters at different temperatures. Figure~\ref{fig:mps}(a) demonstrates that MPS-HEOM method requires a bond dimension of 12 (corresponding to 9184 parameters) to characterize the high-temperature dynamics with $\mathcal{E}_{n_{\uparrow}}=0.81\%$, while the RBMs with 2 hidden and auxiliary nodes (corresponding to 634 parameters) can achieve higher accuracy with $\mathcal{E}_{n_{\uparrow}}=0.07\%$. Similiarly, for low-temperature dynamics (\Fig{fig:mps}(b)), MPS-HEOM requires a bond dimension of 50 (corresponding to 299120 parameters, $\mathcal{E}_{n_{\uparrow}}=0.97\%$) versus an RBM with 15 hidden and auxiliary nodes (corresponding to 5545 parameters, $\mathcal{E}_{n_{\uparrow}}=0.11\%$). From this comparison (summarized in Table~\ref{mps}), we can infer that the representation power and scalability of RBMs for these problems are much superior to those of MPSs. Table~\ref{mps}  further indicates that the number of parameters required in MPS-HEOM significantly exceeds the RDT element count in DQME at low temperature, attributable to the complex non-Markovian environment and the strong system correlations.
%}

%\clearpage

%%%%%%%%%%%%%%%%%%%%%%%%%%%%%%%%%%%%%%%%
%\bibliographystyle{apsrev4-2}%{aip}%
%
\bibliography{ref}
\bibliographystyle{aip}